\documentstyle[emulateapj, psfig]{article}

\received{}
\accepted{}
\journalid{}{}
\articleid{}{}

\slugcomment{Submitted to {\it The Astrophysical Journal Letters}}

\lefthead{Gotthelf {\it et al.}}
\righthead{Pulsed X-ray Emission from PSR B1706$-$44}

\begin{document}

\def\lapp{\ifmmode\stackrel{<}{_{\sim}}\else$\stackrel{<}{_{\sim}}$\fi}
\def\gapp{\ifmmode\stackrel{>}{_{\sim}}\else$\stackrel{>}{_{\sim}}$\fi}
\def\kms{km~s$^{-1}$}
\def\psr{PSR~B1706$-$44}
\def\snr{G343.1$-$2.3}
\def\ro{{\it ROSAT\/}}
\def\asca{{\it ASCA\/}}
\title{Detection of Pulsed X-ray Emission from PSR B1706$-$44}

\author{
E. V. Gotthelf,\altaffilmark{1}
J. P. Halpern,\altaffilmark{1} 
R. Dodson\altaffilmark{2}
}

\altaffiltext{1}{Columbia Astrophysics Laboratory, Columbia University,
550 West 120th Street, New York, NY 10027}

\altaffiltext{2}{School of Mathematics and Physics, University of Tasmania, GPO Box 252--21, Hobart, Tasmania 7001, Australia}

%\altaffiltext{3}{Australia Telescope Natioal Facility, CSIRO, P.O. Box 76, 
%Epping NSW 1710, Australia}

\begin{abstract}

We report the first detection of pulsed X-ray emission from the young,
energetic radio and $\gamma$-ray pulsar \psr.
We find a periodic signal at a frequency of $f = 9.7588088 \pm
0.0000026$~Hz (at epoch 51585.34104 MJD), consistent with the radio
ephemeris, using data obtained withthe High Resolution Camera on-board
the {\it Chandra X-ray Observatory}.  The probability that this
detection is a chance occurrence is $3.5 \times 10^{-5}$ as judged by
the Rayleigh test.  The folded light curve has a broad, single-peaked
profile with a pulsed fraction of $23\% \pm 6\%$. This result is
consistent the \ro\ PSPC upper limit of $<18$\% after allowing for the
ability of {\it Chandra } to resolve the pulsar from a surrounding
synchrotron nebula.
We also fitted {\it Chandra} spectroscopic data
on \psr, which require at least two components, e.g., a
blackbody of $T_{\infty} = (1.66_{-0.15}^{+0.17}) \times 10^6$~K
and a power-law of
$\Gamma = 2.0 \pm 0.5$.  The blackbody radius at the nominal 2.5 kpc
distance is only $R_{\infty} = 3.6 \pm 0.9$~km,
indicating either a hot region on a cooler
surface, or the need for a realistic atmosphere model that would
allow a lower temperature and larger area.  Because the 
power-law and blackbody spectra each contribute more than
$23\%$ of the observed flux, it is not possible to decide which
component is responsible for the modulation in the
spectrally unresolved light curve.

\end{abstract}


\section{Introduction}
\label{sec:intro}

\psr\ is a 102 ms radio pulsar discovered by Johnston et al. (1992)
and possibly associated with the supernova remnant \snr\ 
(McAdam, Osborne, \& Parkinson 1993).  
With characteristic age $P/2\dot P = 17,500$~yr and spin-down
luminosity $\dot{E} = 3.4 \times 10^{36}$~ergs~s$^{-1}$ \psr\ is
considered a young, energetic neutron star. The distance to the pulsar
is somewhate uncertain.  According to the Taylor \& Cordes (1993) free
electron model of the Galaxy the pulsar lies 1.8~kpc away, however
Koribalski et al. (1995) find a kinematic distance in the range
$2.4-3.2$ kpc from H~I absorption.  Originally detected as the
    high-energy $\gamma$-ray source 2CG342$-$02 by the {\it COS B\/}
satellite (Swanenburg et al. 1981), \psr\ is one of approximately
eight rotation-powered pulsars that are responsible for some of the
brightest $\gamma$-ray sources in the sky.
Thompson et al. (1992; 1996) showed that photons detected from the
direction of \psr\ by the EGRET instrument on board the {\it Compton
Gamma-Ray Observatory} are pulsed at the radio period.  All of the
known $\gamma$-ray pulsars are well observed at X-ray energies between
0.1 and 10~keV, where both surface thermal emission and magnetospheric
synchrotron emission can be studied.  However, {\it pulsed\/} X-ray
emission has not been detected from \psr\ despite several searches.
This failure has hindered attempts to understand its X-ray emission
mechanism(s).

Using an observation of \psr\ by the \ro\ PSPC,
Becker, Brazier, Trumper (1995) placed an upper limit
of 18\% on the pulsed fraction.  They attributed this
negative result to the diluting effect of an as-yet unresolved
synchrotron nebula, recalling the history of Vela (\"Ogelman,
Finley, \& Zimmerman 1993),
a pulsar of similar age and spin-down luminosity.
Finley et al. (1998), using data from the \asca\ GIS, placed
an upper limit of 22\% on the pulsed fraction in the 2$-$10~keV
band, although their image was severely contaminated by stray light
from the bright low-mass X-ray binary 4U1705$-$44
just outside the field of view.  Finley
et al. also analyzed a \ro\ HRI observation, from which they concluded
that $57 \pm 12$\% of the photons associated with \psr\ 
actually come from
a compact nebula with an exponential scale length of
$\approx 27^{\prime\prime}$.  When folding only the photons from the
HRI point source at the radio period, they derived an upper
limit of 29\% on its pulsed fraction.
An observation by the non-imaging {\it Rossi X-ray Timing Explorer}
performed during a low state of 4U1705$-$44
also failed to detect pulsations in the $9-18.5$~keV band.
(Ray, Harding, \& Strickman 1999).
Here we report on {\it Chandra} observations of
\psr\ in which we detect its pulsed emission in X-rays.

\section{Observations}
\label{sec:obs}

The field of \psr/\snr\ was observed by {\it Chandra} twice,
once with each of the two types of cameras at
the focal plane of the telescope. Herein we analyze data from these
observations, made available though the {\it Chandra} public archive.

Timing data on \psr\ were acquired on 2000 Feb 11 with the imaging High
Resolution Camera (HRC-I; Murray et al. 1997), a multichannel plate
detector with an effective time resolution degraded to $\sim 4$~ms due
to a known timing error.  The HRC is sensitive to X-rays in the energy
range $0.1-10$ keV;
%over a $0.\!^{\circ}5 \times 0.\!^{\circ}5 $ square field-of-view;
no useful spectral information is available.  The
target was placed near ($\approx 28^{\prime\prime}$) the optical axis
where the telescope mirror point-spread function (PSF) has half
power radius (the radius enclosing 50\% of the total source counts) of
$\sim 0.\!^{\prime\prime}5$ for energies $E < 6$~keV. The HRC-I
oversamples the PSF with pixels $0.\!^{\prime\prime}1318$ on a side,
which allows the pulsar to be isolated from any surrounding emission
such as the pulsar wind nebulae (PWNe) typically found associated with
young pulsars observed by {\it Chandra} (e.g., see Gotthelf 2001).

%Complimentary to the HRC observation,
Data were also obtained
with the Advanced CCD Imaging Spectrometer (ACIS; Burke et al. 1997),
which has resolution
$\Delta E / E \sim 0.1$ at 1 keV scaling as $1 / \sqrt{E}$ over its
0.2--10~keV active band-pass.  The pulsar was observed on 2000 Aug 14
and positioned on the back-illuminated S3 chip of the ACIS-S array,
offset by $0\farcm5$ from the aim-point, where the PSF is
undersampled by the  $0.\!^{\prime\prime}4920 \times
0.\!^{\prime\prime}4920$ CCD pixels.
Data were collected in the
nominal timing mode, with 3.241~s exposures between CCD readouts, and
in ``FAINT'' spectral mode.
%In the following analysis, data from the
%ACIS and HRC instruments are used exclusively for spectroscopy and
%timing analysis, respectively, while images from both sensors are
%examined.

The standard {\it Chandra} screening criteria produced a total usable
exposure time of 46.8~ks and 14.3~ks for the HRC and ACIS data sets,
respectively.  These resulted in a similar number of detected photons
from the source, and the images made from the two instruments are
found to be consistent. The ACIS image, restricted to the $0.3 -
8$ keV energy band to reduce instrumental background, 
%is shown in Figure~\ref{fig:image}]. The image 
%
%
reveals a point-like X-ray source at the radio pulsar position and
four nearby faint X-ray sources that are coincident with USNO stars to
sub-arcsecond accuracy. 
The {\it Chandra} position of \psr\ is $\alpha
= 17^{\rm h} 09^{\rm m} 42^{\rm s}.73$, $\delta = -44^{\circ}
29^{\prime} 08\farcs4$ (J2000.0) with RMS error $0\farcs5$.  This is
in agreement with the precise radio timing position of Wang et
al. (2000, see Table 1).
%$\alpha = 17^{\rm h} 09^{\rm m} 42^{\rm s}.728$, $\delta =
%-44^{\circ} 29^{\prime} 08\farcs24$.  Thus, we conclude that the point
%source is the X-ray counterpart of the radio pulsar.

To look for evidence of extended emission around \psr\ we constructed
a radial profile, the distribution of counts per unit area
in the HRC as a function of
distance from the pulsar.  For comparison, we
analyzed a 23 ks {\it Chandra} observation of the millisecond pulsar
PSR J0437--4715 observed at the aim-point of the HRC just two days later (Zavlin
et al. 2002).  PSR J0437--4715 is an isolated point source with a spectrum
similar to that of \psr\ (as shown in \S4),
thus providing a realistic example of the in-orbit
point-spread function.  Figure~\ref{fig:profile} displays the two
radial profiles after adjusting for different background intensities,
sampled in an annulus of
$30^{\prime\prime} < r < 60^{\prime\prime}$ where the background is
flat.  We then normalized the two profiles to the peak intensity within
$r <1.\!^{\prime\prime}0$.  While the majority of the emission is
point-like, there is clearly diffuse emission out to a radius of
$\approx 20^{\prime\prime}$.  There is also a bump of
enhanced emission between $11^{\prime\prime} < r < 23^{\prime\prime}$.
This result is consistent with the analysis of these data by Dodson \&
Golap (2002), who displayed extended emission out to $\approx
5^{\prime\prime}$, and is marginally consistent with the
decomposition of the \ro\ HRI image deduced by Finley et al. (1998).
In order to make a direct comparison with the \ro\ energy band, we
also extracted photons from the ACIS image in the range $0.2-2.0$ keV.
After background subtraction, we find 513 photons in the ACIS from the
point source (radius $1.\!^{\prime\prime}5$), and 383 nebular photons
in the annulus $1.\!^{\prime\prime}5 < r < 20^{\prime\prime}$.

\section{Timing Analysis}

We searched for a pulsed signal from \psr\ in the HRC data using
time-tagged photon events extracted from an $3.\!^{\prime\prime}0$
radius aperture centered on the X-ray pulsar position.  This aperture
effectively excludes most of the diffuse emission from the putative surrounding
synchrotron nebula; furthermore, less than 3\% of the photons in this
aperture are background events.  The arrival times of the 824 selected
photons were corrected to the solar system barycenter using the
JPL-DE200 ephemeris and the radio timing position.
We then generated a periodogram using the $Z_1^2$
statistic (Rayleigh test) on a range of test frequencies centered on
the predicted pulsar frequency derived from a fit to
radio observations from the Parkes Observatory archives
(Wang et al. 2000) that span the X-ray observation.
The residual deviations of the radio pulse arrival times from
the fitted quadratic ephemeris in Table 1 are less than 1 ms,
including one observation that
occurred 24 days before the X-ray observation.  The known dispersion
measure was used to correct the radio pulse arrival times to
infinite frequency for absolute phase comparison with the X-ray pulse.
%the predicted pulsar frequency as extrapolated from the radio
%ephemeris (epoch MJD $ 51945 - 52235 $) in the ``Australian Pulsar
%Timing Data Archive''\footnote{Available at {\tt
%www.atnf.csiro.au/research/pulsar/archive}} (cf. Wang et al. 2000).

\begin{figure}
\centerline{
\psfig{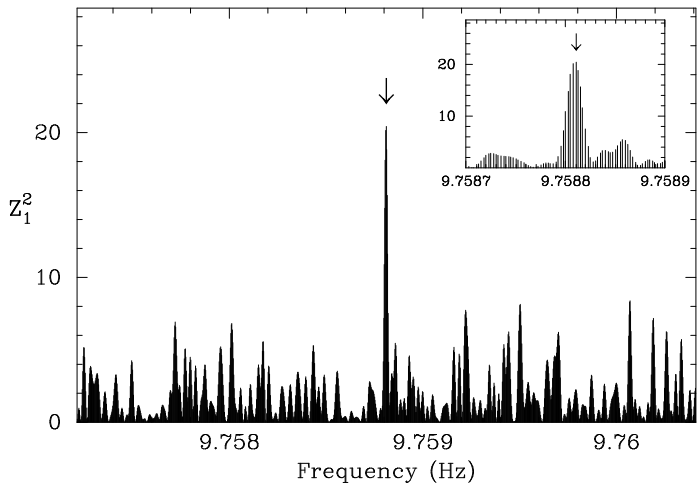}
}
\figcaption{Evidence for pulsed X-ray emission from
\psr\ using a Z$_1^2$ periodogram. The arrow indicates the frequency
expected according to the extrapolated radio ephemeris (see text).
The inset is an expanded view around the peak.
The peak value $Z_1^2 = 20.54$ at $f = 9.7588088$ Hz
has a $3.5 \times 10^{-5}$ probability of
occurring by chance. \label{fig:periodogram}}
\end{figure}

\begin{figure}
\centerline{
\psfig{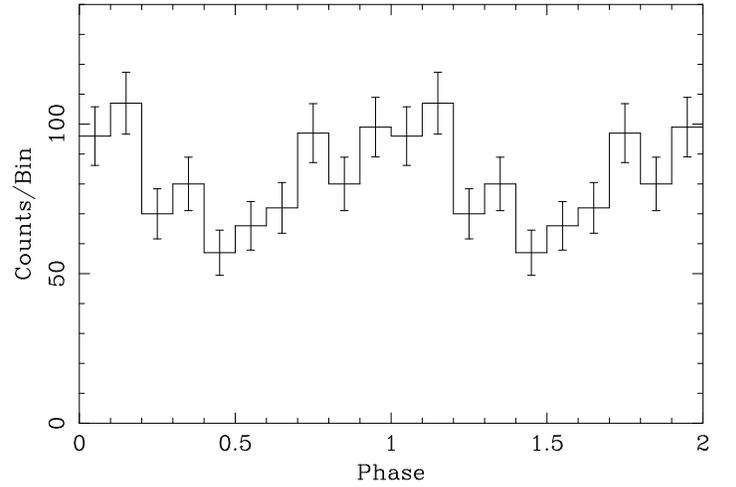}
}
\figcaption{The light curve of \psr\ from the {\it Chandra} HRC
folded according to the radio ephemeris of Table 1.
The quasi-sinusoidal modulation has
pulsed fraction of $23\%\pm 6\%$.
Phase zero corresponds to the epoch of the radio peak.
The error bars are $1 \sigma$.
\label{fig:lightcurve}}
\end{figure}

The X-ray periodogram is shown in Figure~\ref{fig:periodogram}.
We find a peak at $f = 9.7588088 \pm
0.0000026$~Hz (epoch 51585.34104 MJD) which, within the quoted 68\%
uncertainty range, is identical to the contemporaneous radio
frequency, $f = 9.7588097$~Hz.
The $Z_1^2$ statistic for this peak is 20.54, which has a
probability of chance occurrence of $3.5 \times 10^{-5}$.  The light
curve, shown in Figure~\ref{fig:lightcurve}, is broad and
single peaked with a pulsed fraction of $23\% \pm 6\%$.  According to
convention, the pulsed fraction is defined as the ratio of counts
above the minimum in the light curve to the total counts.  This
detection is consistent with the previous upper limits of $<29$\% from
the \ro\ HRI (Finley et al. 1998) and $<18$\% from the {\it ROSAT\/}
PSPC (Becker et al. 1995).  The latter did not resolve the
pulsar from the nebula, therefore, it represents a true upper limit of
$<31\%$ given the contribution of diffuse photons measured in the {\it
Chandra} ACIS image as described in \S 2.
%Based on the statistical
%significance of the HRC signal and its near coincidence with the
%expected period, we regard it as the first detection of X-ray
%pulsations from this well-known radio and gamma-ray pulsar.
The arrival time of the radio pulse is consistent with occurring at
the center of the broad X-ray peak.  This is not the case for
the EGRET $\gamma$-ray pulse, which is centered $\approx 0.37$
cycles after the radio pulse (Thompson et al. 1996).

\section{Spectroscopy}

\begin{figure}
\centerline{
\psfig{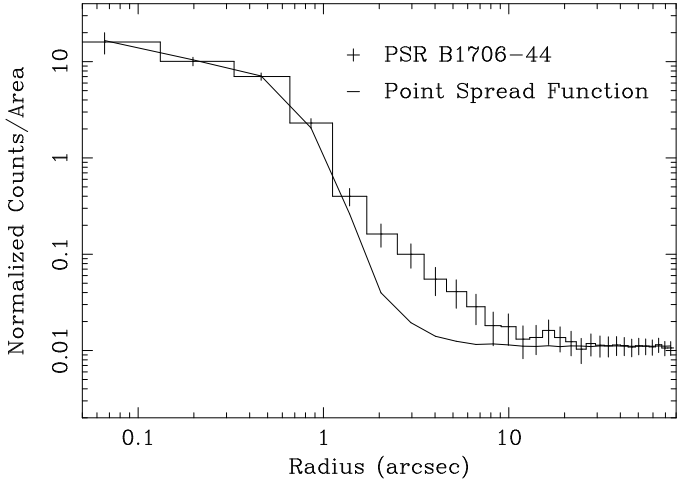}
}
\figcaption{X-ray intensity of \psr\ as a function of radius
measured with the {\it Chandra} HRC ({\it histogram}),
compared with the HRC response to a point source ({\it solid line}; see text). 
The radial profile shows clear evidence for extended emission between
$1.\!^{\prime\prime}5$ and $20^{\prime\prime}$.
The error bars are $1 \sigma$. 
\label{fig:profile}}
\end{figure}

To characterize the energy dependence of the emission from the pulsar
and its putative wind nebula, we analyzed spectral data obtained with
the ACIS detector. Pulsar and PWN source counts spectra were extracted
from two concentric regions, a circle of radius $1^{\prime\prime}$ and
an annulus of radii $2^{\prime\prime}< r <10^{\prime\prime}$,
respectively.  An annular background region centered on the pulsar was
extracted from radii $30^{\prime\prime}< r <60^{\prime\prime}$, which
appears to be pure background (see Fig.~\ref{fig:profile}).  We
verified that background is a negligible contribution to the pulsar
spectrum and a 10\% contribution to the PWN spectrum over their
respective extraction regions. To be specific, there are 561(486)
source counts and an estimated 1(54) background count(s) for the
pulsar(PWN) regions in the $0.5 - 8.0$ keV spectral fitting range.
From the radial profile (Fig. \ref{fig:profile}) it is clear that
there is negligible cross-contamination between the pulsar and PWN
extraction regions.

The ACIS pulse-height data were corrected for gain and resolution
degradation due to CCD charge transfer inefficiencies (CTI) using the
{\it correctit\/} software (Townsley et al. 2001).
%applied to the Level 1 event file.
%We selected the ASCA-like 0,2,3,4,6 grades only and used
%the Townsley et al. 
We used the supplied instrument response matrices appropriate for the CTI
corrected data.  Since the location of the target on the CCD straddled
regions covered by two response files, we generated a
count-weighted mean response to match the data.  The CIAO tool {\tt
mkarf} were used to generate a point-source mirror response matrix for the
pulsar and PWN source regions.  The extracted photons were binned in
pulse-height channel space to match the response matrices and regrouped
such that each fitted spectral bin contained a minimum of 20
counts. The resulting spectra were then fitted with the X-ray spectral
analysis package {\tt XSPEC} (version 11.1) using three different
models, a power law, a simple blackbody, and a sum of both, 
each with interstellar absorption.

\begin{figure}
\psfig{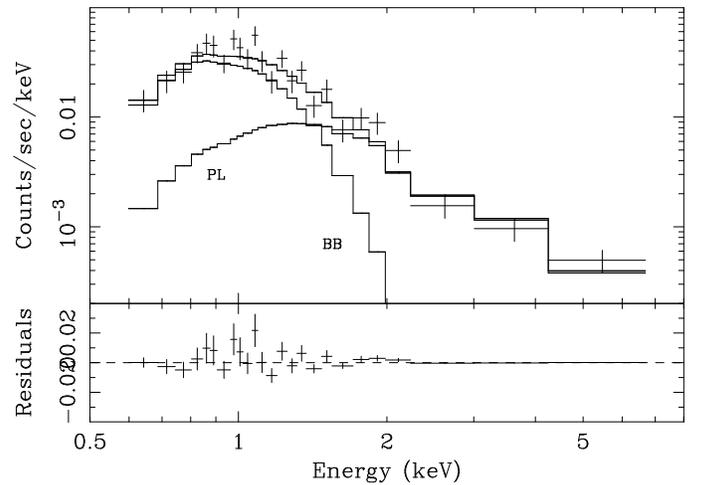}
\figcaption{{\it Chandra} ACIS-S3 spectrum of the point source
coincident with \psr.  {\it Top panel\/:} Data ({\it crosses}) and
best-fit model ({\it dark line}) for the parameters given in
Table~\protect\ref{ta:spectrum}; {\it thin lines} show the contribution 
of the power-law (PL) and blackbody (BB) components of the fit.
{\it Bottom panel\/:} Difference between the data and model, in the
same units as the top panel.\label{fig:spectrum}}
\end{figure}

Initial fits to the pulsar spectrum using blackbody or
power-law models were unacceptable, with $\chi^2 = 115$ and $\chi^2 =
60$ for 23 degrees-of-freedom (DoF), respectively.  We next
determined, as is often the case for young pulsars, that a blackbody
plus power-law fit provides an adequate description of the
spectrum.  The intrinsic parameters of this multicomponent model are
rather unconstrained and strongly correlated with the fitted value of
the interstellar absorption column density, N$_{\rm H}$. To remove an
unnecessary degree of freedom, we fixed $N_{\rm H}$ at the value
obtained from a simple fit to the PWN,
$5.5 \times 10^{21}$~cm$^{-2}$, since we expect the intervening
column density to be the same.  The PWN is well characterized
by a single absorbed power-law with $\chi^2 = 13$ for 20 DoF.
The results of these fits are given in Table~\ref{ta:spectrum},
and the counts spectrum of the pulsar, with the model components 
superposed, are shown in Figure~\ref{fig:spectrum}.

The fitted blackbody component yields effective temperature
$T_{\infty} = (1.66_{-0.15}^{+0.17}) \times 10^6$~K,
bolometric luminosity $L_{\infty} = 6.8_{-1.1}^{+0.8}\times 10^{32}$,
and effective radius $R_{\infty} = 3.6 \pm 0.9$~km in the observed frame.
Here we adopt a distance of 2500~pc as a compromise
between the smaller DM distance of 1.8~kpc, and the H~I kinematic
distance of $2.4-3.2$ kpc.
While $T_{\infty}$ is consistent
with some standard neutron star cooling curves (Umeda, Tsuruta, \& Nomoto
1994), the associated $R_{\infty}$ is significantly smaller
than theoretical neutron star radii.  As has been found frequently for
other pulsars, e.g., Vela (Pavlov et al. 2001), such an outcome could
have either of two implications.  First, the thermal X-rays could be coming
from a smaller hot region on the surface of a cooler
neutron star, the substantial interstellar $N_{\rm H}$ making it difficult
to measure softer X-rays coming from the full surface.  Second, a more
realistic atmosphere model, especially if dominated by hydrogen or helium,
would require a lower effective temperature and larger effective radius,
perhaps becoming consistent with emission from the full neutron star surface.

\section{Discussion and Conclusions}
\label{sec:disc}

There are at least two mechanisms for the emission
of broad-band pulsed X-rays from young rotation-powered pulsars like \psr.  
One is non-thermal magnetospheric synchrotron
from relativistic electrons and positrons created
either in regions above the neutron star polar caps or in
outer gaps.  The second is thermal emission from
the hot surface, a result of initial cooling of the hot neutron star
or reheating of the polar caps by back-flowing accelerated particles.
Sometimes the shape of the pulse sheds additional light
on the X-ray emission mechanism.  Sharp, narrow pulses of high
amplitude can only be produced by a highly beamed, thus relativistic
population of electrons, while quasi-sinusoidal pulses
of low amplitude such as describe \psr\ can be produced by either mechanism.  
For those intermediate-age pulsars ($10^4 - 10^6$~yr) that are also
EGRET sources, the presence of both types of X-ray source
is usually discovered when spectrally resolved timing data are available
(e.g., Wang et al. 1998; Pavlov et al. 2001). 
Unfortunately, in this case, the {\it Chandra} HRC has little or
no energy resolution.  Since each spectral component fitted
to the ACIS spectrum contributes more than 23\% of the flux in
the total energy band to which the HRC is sensitive (60\% from blackbody,
40\% from power law), the pulsed fraction alone does not reveal the
source of the pulsed X-rays.  Either or both components
may contribute to the modulation.
%Additional clues,
%such as the absolute phase relationship of the X-ray pulse
%to the radio and $\gamma$-ray peaks cannot be examined yet
%due to our lack of a closely contemporaneous radio observation
%of this noisy, glitching pulsar (Johnston et al. 1995; Wang et al. 2000).
Future spectrally
resolved X-ray observations with high throughput and
moderate time resolution, such as with {\it XMM-Newton},
could resolving this ambiguity.

The $>100$~MeV luminosity of most $\gamma$-ray pulsars is
a significant fraction of their spin-down power.  In the case
of \psr\, this fraction is $\approx 0.20$ if isotropic, while
its $0.5-8$~keV non-thermal X-ray luminosity is only
$1.3 \times 10^{-4}\,\dot E$ including both pulsar and nebula,
similar to that of other pulsars.
Thompson et al. (1996) parameterized the EGRET spectrum
of \psr\ as a broken power law, with photon index $\Gamma = 1.27 \pm 0.09$
from 50 MeV to 1~GeV, steepening to $\Gamma = 2.25 \pm 0.13$ above 1~GeV.
Since there is no evidence for any {\it unpulsed} $\gamma$-ray emission,
the EGRET spectrum can be compared directly with the power-law
component of the pulsar point source in the {\it Chandra} ACIS
spectrum.  When extrapolated back to 10 keV, the
EGRET spectrum nearly matches the X-ray flux, although the
power-law X-ray slope itself, with $\Gamma = 2.0 \pm 0.5$,
is somewhat steeper than the EGRET value.
The non-thermal X-rays are not likely to be emitted
by the same population of electrons/positrons as produce the
$\gamma$-rays, but may instead originate from a much
less energetic population, such as secondary pairs from
the inward emitted $\gamma$-rays converting on the strong $B$-field
near the neutron star surface.  This distinction is also supported
by the observed phase offset between the X-ray and $\gamma$-ray
pulses (\S 3).  The Wang et al. (1998) 
model predicts $L_x$ (2--10~keV)
$\approx 2 \times 10^{31}$ ergs~s$^{-1}$
and $\Gamma = 1.5$ from this process for \psr, which is
similar to its observed
non-thermal X-ray component.  If its {\it thermal\/} luminosity
comes partly from a small surface area, then inward 
flowing primary electrons impacting the polar caps
may be responsible.  Wang et al. predicted
$L_x \approx 1 \times 10^{33}$ ergs~s$^{-1}$ of thermal 
emission from this process,
not far from the observered thermal luminosity of \psr.

%e.g. \cite{dh82} or in outer gaps, e.g. \cite{ry95}.  The
%measured power-law photon index (Table~\ref{ta:spectrum}) is
%consistent with that observed for pulsed blackbody emission from other
%rotation-powered pulsars (\cite{bt97}).  Furthermore, the observed
%2--10~keV flux (Table~\ref{ta:spectrum}) implies an unabsorbed
%luminosity, for a distance $d=1.8$~kpc (ref), of $x \times
%10^{33}$~erg~s$^{-1}$, assuming beaming angle $\phi = \pi$~sr.  This
%implies an efficiency of conversion of spin-down luminosity into
%magnetospheric emission of $y(\phi/\pi \; {\rm sr})(d/2.8 \; {\rm
%kpc})^2$.  This is consistent with that observed for the
%magnetospheric components of other radio pulsars \cite{bt97}.

\acknowledgements 

This work made use of data obtained from the {\it Chandra} Data Archive,
and is supported by NASA LTSA grant NAG 5-7935 to EVG.  We thank D. Lewis
and R. N. Manchester for their assistance with the radio ephemeris.

%\bibliographystyle{apj}
%\bibliography{journals1,modrefs,psrrefs,crossrefs}

%\hbox{}
\twocolumn
%\clearpage

%\hbox
{
\begin{deluxetable}{lr}
\tablenum{1}
\tablecolumns{2}
\tablewidth{0pc}
\tablecaption{Radio Ephemeris of \psr\ }
\tablehead
{
\colhead{Parameter} & \colhead{Value}
}
\startdata
R.A. (J2000)          & $17^{\rm h}09^{\rm m}42.\!^{\rm s}728$         \\
Decl. (J2000)         & $-44^{\circ}29^{\prime}08.\!^{\prime\prime}24$ \\
Valid Range (MJD)			   & $51520-51956$ \\
Epoch (MJD)\tablenotemark{a}       & 51585.3410392671                  \\
Frequency (Hz)                           & 9.7588096779                \\
Frequency Derivative (Hz s$^{-1}$)       & $-8.880796 \times 10^{-12}$  \\
Dispersion Measure (cm$^{-3}$ pc)        & 75.69                       \\
\enddata
\footnotesize
\tablenotetext{a}{TDB epoch of radio peak at infinite frequency.}
\end{deluxetable}
}
\vfill

\begin{deluxetable}{lll}
\tablenum{2}
\tablewidth{280pt}
\tablecaption{Spectral Fits to \psr\ and its PWN\tablenotemark{a}\label{ta:spectrum} }
%\footnotesize
\tablehead{
\colhead{} & \colhead{Pulsar} & \colhead{Nebula}} 
\startdata
$N_{\rm H}$ ($10^{21}$~cm$^{-2}$)     & 5.5 (fixed)      & $5.5_{-2.1}^{+2.5}$    \\
$\Gamma$                        & 2.0$_{-0.5}^{+0.5}$ &    $1.34_{-0.30}^{+0.24}$ \\
$C_{\rm pl}$\tablenotemark{b}   & $4.3_{-1.6}^{+3.1} \times 10^{-5}$
   & $5.5_{-1.8}^{+2.6} \times 10^{-5}$          \\
$F_{\rm pl}$(absorbed)\tablenotemark{c}
                                & $1.2 \times 10^{-13}$  & $3.4 \times 10^{-13}$  \\
$L_{\rm pl}$(unabsorbed)\tablenotemark{d}
                                & $1.45_{-0.08}^{+0.46} \times 10^{32}$   & $3.1_{-1.0}^{+1.5} \times 10^{32}$  \\
$T_{\infty}$ (K)\tablenotemark{e} & $1.66_{-0.15}^{+0.17} \times 10^6$  &  \dots                 \\
$R_{\infty}$ (km)\tablenotemark{e} & $3.6 \pm {0.9}$ &  \dots               \\
$L_{\infty}$ (ergs~s$^{-1}$)\tablenotemark{e} & $6.8_{-1.1}^{+0.8}\times 10^{32}$  &  \dots \\
$\chi^2_{\nu}$ [DoF]           & 1.4 [21]            & 0.61 [21]              \\
\enddata
\footnotesize
\tablenotetext{a}{Uncertainties are 68\% confidence for two interesting parameters.}
\tablenotetext{b}{Power-law normalization in units of
photons~cm$^{-2}$~s$^{-1}$~keV$^{-1}$ at 1~keV.}
\tablenotetext{c}{Power-law flux (ergs~cm$^{-2}$~s$^{-1}$) in the 0.5--8 keV band.}
\tablenotetext{d}{Power-law luminosity (ergs~s$^{-1}$) in the 0.5--8 keV band
for $d = 2500$ pc.}
\tablenotetext{e}{Spherical blackbody parameters for $d = 2500$ pc.}
\end{deluxetable}

\vfill

\clearpage
\twocolumn

\end{document}